\documentclass[12pt]{article}
\usepackage{amsmath,amssymb,epsfig}
\usepackage{cancel}

\makeatletter

\@addtoreset{equation}{section}
\makeatother

\begin{document}

\title{\vbox{
\baselineskip 14pt
\hfill \hbox{\normalsize KUNS-2409}
} \vskip 1.7cm
\bf Massive modes in magnetized brane models \vskip 0.5cm
}
\author{
Yuta~Hamada \ and \
Tatsuo~Kobayashi
\\*[20pt]
{\it \normalsize 
Department of Physics, Kyoto University, 
Kyoto 606-8502, Japan}
 \\*[50pt]}

\date{
\centerline{\small \bf Abstract}
\begin{minipage}{0.9\linewidth}
\medskip 
\medskip 
\small
We study higher dimensional models with magnetic fluxes, 
which can be derived from superstring theory.
We study 
mass spectrum and wavefunctions of 
massless and massive modes for spinor, scalar and vector fields.
We compute the 3-point couplings and higher order couplings 
among massless modes and  massive modes 
in 4D low-energy effective field theory.
These couplings have non-trivial behaviors, because 
wavefunctions of massless and massive modes are non-trivial.
\end{minipage}
}

\newpage

\begin{titlepage}
\maketitle
\thispagestyle{empty}
\clearpage
\end{titlepage}

\renewcommand{\thefootnote}{\arabic{footnote}}
\setcounter{footnote}{0}

\section{Introduction}

Field theory in higher dimensions plays a role in 
particle physics and cosmology.
In particular, extra dimensional field theory derived from 
superstring theory is important.
Their four-dimensional (4D) low-energy effective field theories 
are determined by geometrical aspects of compact extra dimensions.
One of the simplest compact spaces is a torus.
However, the simple toroidal compactification 
does not lead to a chiral theory as 
a 4D low-energy effective field theory.
Hence, it is a key issue to realize a 4D chiral theory when 
we start with higher dimensional field theory.

Complicated geometrical backgrounds such as Calabi-Yau 
manifolds would lead to 
a 4D chiral theory, although it may be difficult to 
compute explicitly 4D low-energy effective field theories from 
such geometrical backgrounds.
On the other hand, the toroidal compactification 
can also lead to a 4D chiral theory when we introduce 
non-vanishing magnetic fluxes in extra dimensions.
The number of zero-modes are determined by 
the size of magnetic flux and each zero-mode has
a quasi-localized profile.
Thus, the toroidal compactification with 
magnetic fluxes is quite attractive background 
for higher dimensional field theory \cite{Cremades:2004wa,Abe:2008fi,
DiVecchia:2008tm,Choi:2009pv,Abe:2012ya} 
(see also \cite{Conlon:2008qi,
Marchesano:2008rg}).
Its stringy setup corresponds to magnetized D-brane models 
wrapping cycles on the torus \cite{Bachas:1995ik,Berkooz:1996km,
Blumenhagen:2000wh,Angelantonj:2000hi}.
Furthermore, magnetized D-brane models are the T-dual of 
intersecting D-brane models, and 
many interesting models have been constructed in both 
types of models \cite{Ibanez,Blumenhagen:2006ci}.

The Yukawa couplings among massless modes were computed 
by integrating the overlap of wavefunctions in the extra dimensional 
space  
\cite{Cremades:2004wa}.
If zero-modes are quasi-legalized far away from each other, 
their couplings are suppressed.
On the other hand, if they are localized near each other, 
their couplings are not suppressed, but would be of ${\cal O}(1)$.
Thus, these localization behaviors are important from the 
phenomenological viewpoint, for example, to derive 
the realistic values of quark and lepton masses and 
their mixing angles.
Furthermore, higher order couplings among massless modes were also 
computed \cite{Abe:2009dr}.
Interestingly, they are written by 
products of 3-point couplings.
These low-energy effective field theories can also lead to 
Abelian and non-Abelian discrete flavor symmetries, e.g. 
$D_4$ and $\Delta(27)$ flavor symmetries \cite{Abe:2009vi,BerasaluceGonzalez:2012vb}.\footnote{
Similar non-Abelian discrete flavor symmetries are derived in 
heterotic orbifold models \cite{Kobayashi:2004ya}.}
These non-Abelian discrete flavor symmetries are important to 
derive the realistic quark and lepton mass matrices
(see e.g. \cite{Ishimori:2010au} and references therein).

In addition to massless modes, massive modes also 
have important effects in 4D low-energy effective field theory.
For example, they may induce the fast proton decay and 
flavor changing neutral currents (FCNCs) (see e.g. \cite{Camara:2011nj}).
Our purpose in this paper is to study massive modes 
in the extra dimensional models with magnetic fluxes.
We study their mass spectrum and wavefunctions explicitly.
Then, we study compute 3-point couplings and higher order 
couplings including these massive modes.
These couplings have non-trivial behaviors, because 
wavefunctions of massless and massive modes are non-trivial.

This paper is organized as follows.
In section 2, we briefly review on the fermion zero-modes 
on $T^2$ with the magnetic flux.
Then, we study mass spectrum and wavefunctions of 
higher modes explicitly.
These analysis is extended to those for zero-modes and higher modes 
of scalar and vector fields.
Its extension to $T^6$ is straightforward.
In section 3, we compute couplings among these modes.
In section \ref{eq:coupling-zero-model}, 
we give a brief review on computations of 
the 3-point couplings and higher order couplings 
among zero-modes.
Then, we extend them to the computations of 
the 3-point and higher order couplings including 
higher modes in \ref{eq:coupling-massive-model}.
In section \ref{eq:coupling-WL}, 
we also consider the couplings including massive modes due to 
only the Wilson line effect, but not magnetic fluxes.
In section 4, we give comments on some phenomenological 
implications of our results.
Section 5 is devoted to conclusion and discussion.
In Appendix A, we show some useful properties of 
the Hermite function.
In Appendix B, we briefly review on the vector field 
in extra dimensions.
In Appendix C, we show useful properties of 
the products of zero-mode wavefunctions.

\section{Mass spectrum and wavefunctions of 
massive modes}


We consider the $(4+d)$-dimensions, and 
denote four-dimensional and $d$-dimensional coordinates 
by $x^\mu$ and $y^m$ with $\mu=0,\cdots,3$ and 
$m=1,\cdots, d$, respectively.
We study the spinor field $\lambda (x^\mu,y^m)$ and 
the vector field $A_M (x^\mu,y^m)$ with $M=0,\cdots,(3+d)$.
We decompose these fields as follows,
\begin{eqnarray}\label{eq:KK-decompose}
\lambda (x^\mu,y^m) &=& \sum_n \chi_n(x^\mu) \psi_n(y^m),  \\
A_M (x^\mu,y^m) &=& \sum_n \varphi_{n,M}(x^\mu) \phi_{n,M}(y^m).
\end{eqnarray}
Here we choose the internal wavefunctions $\psi_n(y^m)$ as 
eigenfunctions of the internal Dirac operator as 
\begin{eqnarray} \label{eq:eigen-f-spinor}
i\Gamma^m D_m \psi_n = m_n \psi_n,
\end{eqnarray}
where $\Gamma^m$ denote the gamma matrices in the internal space.
The eigenvalues of $m_n$ become masses of the modes 
$\chi_n(x^\mu)$ in 4D effective field theory.
Similarly, $\phi_{n,M}(y^m)$ correspond to 
eigenfunctions of the internal Laplace operators, 
as will be shown explicitly later.
The scalar field in the $(4+d)$-dimensions is 
also decomposed in a similar way.

\subsection{$T^2$ with magnetic flux}
\label{sec:T2}

First, let us consider the 2D torus, $T^2$.
Here, we follow the notation of Ref.~\cite{Cremades:2004wa}.
Instead of the real coordinates $y^1$ and $y^2$, 
we use the complex coordinate, $z=y^1+\tau y^2$ with 
$\tau \in {\bf C}$.
The metric is given by 
\begin{eqnarray}
ds^2 = 2(2\pi R)^2 dz d \bar z .
\end{eqnarray}
We identify the complex coordinate 
as $z \sim z +1$ and $z \sim z + \tau$ on $T^2$.
The area is written by ${\cal A}=4\pi^2 R^2 {\rm{Im} \tau}$.

We introduce the U(1) magnetic flux on $T^2$ as 
\begin{eqnarray}
F_{z\bar z} = \frac{\pi i}{\rm{Im} \tau}m.
\end{eqnarray}
This magnetic flux is derived e.g. from the following 
vector potential,
\begin{eqnarray}
A_{\bar z} = \frac{\pi}{2\rm{Im} \tau}m z , \qquad 
A_{z} = - \frac{\pi}{2\rm{Im} \tau}m \bar z .
\end{eqnarray}
Their boundary conditions can be written as
\begin{eqnarray}
A_i(z+1)=A_i(z)+\partial_i \chi_1, \qquad 
A_i(z+\tau)=A_i(z)+\partial_i \chi_2,
\end{eqnarray}
where 
\begin{eqnarray}
\chi_1=\frac{\pi}{{\rm Im}\tau} m~{\rm Im} z, 
\qquad 
\chi_2=\frac{\pi}{{\rm Im}\tau} m~{\rm Im} \bar \tau z.
\end{eqnarray}
Furthermore, 
we can introduce non-vanishing Wilson lines by using 
\begin{eqnarray}
\chi_1=\frac{\pi}{{\rm Im}\tau} {\rm Im} (mz +\alpha), 
\qquad 
\chi_2=\frac{\pi}{{\rm Im}\tau} {\rm Im} \bar \tau (mz + \alpha),
\end{eqnarray}
where $\alpha$ is complex and corresponds to the degree of 
freedom of the Wilson line.
It is convenient to use the following notation,
\begin{eqnarray}
\alpha = m  \zeta,
\end{eqnarray}
for $m\neq 0$.

\subsubsection{Fermion zero-modes}

Here, we review the fermion zero-modes, which satisfy 
Eq.(\ref{eq:eigen-f-spinor}) with $m_n=0$ \cite{Cremades:2004wa}.
On $T^2$, the spinor $\psi_n$ has two components, 
\begin{eqnarray}
\psi_n = 
\begin{pmatrix}
\psi_{+,n} \\
\psi_{-,n}
\end{pmatrix}.
\end{eqnarray} 
We use the gamma matrices on $T^2$ as 
\begin{eqnarray}
\Gamma^1=
\begin{pmatrix} 0 & 1 \\
1 & 0
\end{pmatrix}, \qquad 
\Gamma^2 = 
\begin{pmatrix} 0 & -i \\ i & 0
\end{pmatrix}.
\end{eqnarray}
Then, the zero-mode equation is written as 
\begin{eqnarray}
D\psi_{+,0}= 0, \qquad D^\dagger \psi_{-,0} =0,
\end{eqnarray}
where 
\begin{eqnarray}
D = \frac{1}{\pi R}\left( \bar \partial 
+ q \frac{\pi m}{2 {\rm Im} \tau} (z +  \zeta)
\right),
\end{eqnarray}
for the spinor with $U(1)$ charge $q$.
The charge $q$ and magnetic flux $m$ should satisfy that 
$qm=$ integer.
They also satisfy the following boundary conditions,
\begin{eqnarray}\label{eq:fermion-bc}
\psi_n(z+1) = e^{iq\chi_1(z)}\psi_n(z), \qquad 
\psi_n(z+\tau) = e^{iq\chi_2(z)}\psi_n(z).
\end{eqnarray}

When $qm >0$, only the zero-mode $\psi_{+,0}$ has a solution, but 
$\psi_{-,0}$ has no solution.
Then, the chiral spectrum for the zero-modes is realized and 
the number of zero-modes is equal to $qm$.
Their zero-mode wavefunctions are written explicitly as 
\begin{eqnarray}\label{eq:fermion-zero}
\psi^{j,qm}_+(z+\zeta) = \left( \frac{2 {\rm Im} \tau qm}{{\cal
      A}^2}\right)^{1/4} \sum_\ell \Theta^{j,qm}_{\ell}(z+\zeta,\tau),
\end{eqnarray}
where 
\begin{eqnarray}
& & \Theta^{j,qm}_{\ell}(z+\zeta,\tau) = 
\exp\bigg[-\pi qm \mathrm{Im}\tau
\bigg(\frac{\mathrm{Im}(z+ \zeta)}{\mathrm{Im}\tau}+\frac{j}{qm}+\ell\bigg)^2
  \\   & &~~~~~~~~~~
+i\pi qm\mathrm{Re}(z+\zeta)\bigg(\frac{\mathrm{Im}(z+ \zeta)}{\mathrm{Im}\tau}+2\bigg(\frac{j}{qm}+\ell\bigg)\bigg)
+i\pi qm\mathrm{Re}\tau \bigg(\frac{j}{qm}+\ell \bigg)\bigg]. \nonumber
\end{eqnarray}
Note that the effect of the Wilson line $\zeta$ is the shift of 
the wavefunctions $\psi^{j,qm}(z)$ to $\psi^{j,qm}(z+\zeta)$.
The zero-mode wavefunction can be written by 
a product of the Gaussian function and the Jacobi
$\vartheta$-function, i.e., 
\begin{eqnarray}
\psi^{j,qm}_+(z+\zeta) &=& \left( \frac{2 {\rm Im} \tau qm}{{\cal
      A}^2}\right)^{1/4}\exp \left[ i\pi 
\frac{qm(z+\zeta){\rm Im}(z+\zeta)}{{\rm Im } \tau} \right] \nonumber \\
& & \times \vartheta  \left[ \begin{matrix} j/qm \\ 0
\end{matrix} \right](qm(z+\zeta),qm\tau),
\end{eqnarray}
where 
\begin{eqnarray}
\vartheta \left[ \begin{matrix} a \\ b \end{matrix} \right]
(\nu,\tau)=\sum_\ell \exp \left[ \pi i (a + \ell)^2\tau 
+ 2 \pi i (a + \ell) (\nu +b) 
\right].
\end{eqnarray}
$(\psi^{j,qm}_+)^*$ represents the anti-particle of 
$\psi^{j,qm}_+$, and 
 is obtained from Eq.(\ref{eq:fermion-zero}) 
by replacing $\Theta^{j,qm}_{\ell}(z+\zeta,\tau)$
with $\Theta^{-j,-qm}_{\ell}(\bar z + \bar \zeta,\bar \tau)$.
These zero-mode wavefunctions satisfy the following 
orthonormal condition,
\begin{eqnarray}\label{eq:orthonormal}
\int_{T^2}dz d\bar z \psi^{j,qm}_+ (\psi^{k,qm}_+)^* = \delta_{jk}.
\end{eqnarray}

When $qm < 0$, there appear the zero-modes for $\psi_{-,0}$, 
but not for $\psi_{+,0}$.
The number of their zero-modes is equal to $|qm|$, and 
their wavefunctions are obtained similarly.
In the following discussions, we assume $qm >0$.

\subsubsection{Fermion massive modes}

Here, we study the fermion massive modes with $m_n \neq 0$ 
in Eq.~(\ref{eq:eigen-f-spinor}).
For $m_n \neq 0$, the zero-modes, $\psi_{+,n}$ and 
$\psi_{-,n}$, mix each other in Eq.~(\ref{eq:eigen-f-spinor}).
They satisfy 
\begin{eqnarray}
\begin{pmatrix} 
D^\dagger D & 0 \\  0 & D D^\dagger
\end{pmatrix} 
\begin{pmatrix}
\psi_{+,n} \\ \psi_{-,n}
\end{pmatrix} = m^2_n 
\begin{pmatrix}
\psi_{+,n} \\ \psi_{-,n}
\end{pmatrix} .
\end{eqnarray}

The 2D Laplace operator is defined as 
\begin{eqnarray}
\Delta = \frac12 \{ D^\dagger, D \},
\end{eqnarray}
and it satisfies the following algebraic relations,
\begin{eqnarray}
& & \Delta = D^\dagger D + \frac{2\pi qm}{\cal A}, 
\qquad [D,D^\dagger]= \frac{4 \pi qm}{\cal A}, \nonumber \\
& & [\Delta, D^\dagger]=\frac{4 \pi qm}{\cal A}D^\dagger, \qquad 
 [\Delta, D]=-\frac{4 \pi qm}{\cal A}D.
\end{eqnarray}
Thus, massive modes are eigenfunctions of the Laplace operator 
$\Delta$, and their mass spectrum is derived in an analysis 
similar to the quantum harmonic oscillator.
It is convenient to use the normalized 
creation and annihilation operators, 
\begin{eqnarray}
a=\sqrt{\frac{\cal A}{4\pi qm} }D, \qquad
a^\dagger=\sqrt{\frac{\cal A}{4\pi qm} }D^\dagger, 
\end{eqnarray}
which satisfy $[a,a^\dagger]=1$.
Then, the eigenvalues of the Laplace operator $\Delta$ are 
given as 
\begin{eqnarray}
\lambda_n = 2 \pi \frac{qm}{\cal A} (2n+1),
\end{eqnarray}
and eigenvalues $m_n^2$ are also written as 
 \begin{eqnarray}
m_n^2 = 4 \pi \frac{qm}{\cal A} n .
\end{eqnarray}
The corresponding wavefunctions $\psi_n$ are written by
\begin{eqnarray}
\frac{1}{\sqrt n!}(a^\dagger)^n \psi^{j.qm}_{+,0}.
\end{eqnarray}
Explicitly, the wavefunctions of massive modes are written as 
\begin{eqnarray}\label{eq:massive-wf}
\psi_n^{j,qm}&=&\frac{{(2mq\mathrm{Im}\tau})^{1/4}}{(2^n n! {\cal A})^{1/2}}
\sum_l \Theta^{j,qm}_{\ell}(z+\zeta,\tau) \nonumber \\
& & \times 
H_n\bigg(\sqrt{2\pi qm \mathrm{Im}\tau}\bigg(\frac{\mathrm{Im}(z+\zeta)}{\mathrm{Im}\tau}
+\frac{j}{qm}+\ell\bigg)\bigg),
\end{eqnarray}
where $H_n(x)$ is the Hermite function.
Massive spectra of $\psi_{+,n}$ and $\psi_{-,n}$ are the
same and the number of each of them is equal to $qm$.
Note that $\psi_n^{j,qm}$ satisfy the boundary conditions 
(\ref{eq:fermion-bc}).
Also, these wavefunctions satisfy the following orthonormal conditions,
\begin{eqnarray}
\int_{T^2}dz d\bar z \psi^{j,qm}_n(\psi^{k,qm}_\ell)^* = \delta_{jk}
\delta_{n \ell}.
\end{eqnarray}

\subsubsection{Scalar and vector modes}

Here, we study the scalar and vector modes on $T^2$.
The scalar fields are expanded as eigenfunctions of 
the Laplace operator,
\begin{eqnarray}
\Delta \phi_n (z) = m^2_n\phi_n (z).
\end{eqnarray}
That is, these eigenvalues are obtained as $\lambda_n$,
i.e.,
\begin{eqnarray}
m^2_n = \lambda_n = 2 \pi \frac{qm}{\cal A} (2n+1),
\end{eqnarray}
for the scalar field.
All of them including the lightest mode with $n=0$ are massive.
Eigenfunctions are the same as those for the fermion, i.e. 
$\psi_{n}^{j,qm}$ in Eq.(\ref{eq:massive-wf}).

Next, we study the vector field on $T^2$.
We are interested in the charged vector field 
with the $U(1)$ charge $q$, where $q\neq 0$.\footnote{
Obviously, there is no effect due to magnetic fluxes 
in the neutral vector fields with $q=0$.}
For example, they correspond to the 
$W^\pm$ vector bosons in the $SU(2)$ gauge theory.
We decompose the vector fields as Eq.~(\ref{eq:KK-decompose}).
{}From Eq.~(\ref{eq:vector-mass}) in Appendix \ref{sec:app-vec}, 
the mass-squared matrix is written by 
\begin{eqnarray}
{\cal M}^2=
\begin{pmatrix}
\Delta & -i 4 \pi \frac{qm}{\cal A} \\
i 4 \pi \frac{qm}{\cal A} & \Delta 
\end{pmatrix},
\end{eqnarray}
in the real basis of the 2D vector field, $(\phi_{n,1},\phi_{n,2})$.
Instead of the real basis, we use the complex basis, 
\begin{eqnarray}
\phi_{n,z}=\frac{1}{\sqrt 2}(\phi_{n,1} + i \phi_{n,2}), \qquad 
\phi_{n,\bar z}=\frac{1}{\sqrt 2}(\phi_{n,1} - i \phi_{n,2}).
\end{eqnarray}
The mass spectra of these internal wavefunctions are 
obtained through solving the following equations,
\begin{eqnarray}
 & &  
\left( \Delta - \frac{4\pi qm}{\cal A} \right) \phi_{n,z}=m_n^2\phi_{n,z},  \\
& & \left( \Delta + \frac{4\pi qm}{\cal A} \right)
\phi_{n,\bar z}=m_n^2\phi_{n,\bar z} .
\end{eqnarray}
That is, the mass spectrum of $\phi_{n,z}$ is obtained as 
\begin{eqnarray}
m^2_n = \lambda_n- \frac{4\pi qm}{\cal A} = 2 \pi \frac{qm}{\cal A} (2n-1),
\end{eqnarray}
while the mass spectrum of $\phi_{n,\bar z}$ is obtained as 
\begin{eqnarray}
m^2_n = \lambda_n+ \frac{4\pi qm}{\cal A} = 2 \pi \frac{qm}{\cal A} (2(n+1)+1).
\end{eqnarray}
The spectrum of $\phi_{n,z}$ includes the tachyonic mode for $n=0$, 
while all modes of $\phi_{n,\bar z}$ are massive.
Their wavefunctions, $\phi_{n, z}$ and $\phi_{n,\bar z}$, are 
the same as those for the fermion, i.e. $\psi^{j,qm}_n$  in Eq.(\ref{eq:massive-wf}).

\subsubsection{Massive modes only due to Wilson lines}

The massive modes also appear only due to non-vanishing Wilson lines 
$\alpha$ without magnetic fluxes.
For completeness, we show their mass spectrum and wavefunctions.
The internal wavefunctions for the spinor field as well as 
the scalar and vector fields satisfy the 
same boundary condition as Eq.~(\ref{eq:fermion-bc}) 
with 
\begin{eqnarray}
\chi_1 = \frac{\pi}{{\rm Im}\tau}{\rm Im} \alpha, \qquad 
\chi_2 = \frac{\pi}{{\rm Im}\tau}{\rm Im} \bar \tau \alpha.
\end{eqnarray}
Then, the wavefunctions satisfying this boundary condition are obtained as 
\begin{eqnarray}\label{eq:massive-WL}
\psi_{n_R,n_I}^{(W)}(z) = \frac{1}{\sqrt {\cal A}}\exp \left[
i\pi \left( \frac{{\rm Im}\alpha }{{\rm Im}\tau} + 2n_R \right) {\rm Re} z + 
i \pi \frac{{\rm Im }z}{{\rm Im} \tau} \left(-{\rm Re}\alpha + 
2 (n_I - {\rm Re} \tau n_R) \right) \right], 
\end{eqnarray}
where $n_R, n_I$ are integers.
Their masses are given as 
\begin{eqnarray}
m^2_{n_R,n_I} &=& \frac{4\pi^2{\rm Im}\tau}{{\cal A}}
\bigg[ \left(  \frac{{\rm Im}\alpha}{{\rm Im}\tau} + n_R
\right)^2 \nonumber \\ 
& & 
+ \left(\frac{1}{{\rm Im} \tau} \right)^2\left(-{\rm Re}\alpha + 
 (n_I - {\rm Re} \tau n_R) \right)^2 \bigg].
\end{eqnarray}

\subsection{$T^6$}

Here we study the field theory on $(T^2)^3$.
It is straightforward to extend the analyses on $T^2$ and $(T^2)^3$ to 
one on  $(T^2)^2$.
We use the complex basis, $z^i=y^{2i-1}+\tau^i y^{2i}$ with $i=1,2,3$ 
on the $i$-th $T^2$, 
and the metric is written by 
\begin{eqnarray}
ds^2= \sum_i 2(2\pi R^i)^2dz^i d \bar z^i.
\end{eqnarray}
We identify the complex coordinate 
as $z^i \sim z^i +1$ and $z^i \sim z^i + \tau^i$, and 
the area on the $i$-th $T^2$ is written by 
${\cal A}^i=4\pi^2 (R^i)^2 {\rm{Im} \tau^i}$.

We introduce the U(1) magnetic flux on the $i$-th $T^2$ as 
\begin{eqnarray}
F_{z^i\bar z^i} = \frac{\pi i}{\rm{Im} \tau^i}m^i,
\end{eqnarray}
where $qm^i$ is integer.
This magnetic flux is derived from the following 
vector potential,
\begin{eqnarray}
A_{\bar z^i} = \frac{\pi}{2\rm{Im} \tau^i}m^i z^i , \qquad 
A_{z^i} = - \frac{\pi}{2\rm{Im} \tau^i}m^i \bar z^i .
\end{eqnarray}
We also introduce the Wilson line on the $i$-th $T^2$, 
\begin{eqnarray}
\alpha^i =m^i\zeta^i.
\end{eqnarray}

Obviously, the mass spectrum and wavefunctions on each $T^2$ 
are given as those in section \ref{sec:T2}.
The full eigenfunctions are the products of  
the eigenfunctions for the $n_i$-th modes on the $i$-th 
$T^2$, and the full mass squared is the sum of masses 
squared for each $T^2$ .
The number of massless fermions are obtained as 
$\prod_i qm^i$.
The scalar field on $T^6$ is always massive.
The vector field $\phi_{z^r}$ along the $r$-th (complex) direction 
on $T^6$ has the lowest mass squared with $n^i=0$ $(i=1,2,3)$ as 
\begin{eqnarray}
m^2= 2\pi q \left( \sum_{i\neq r} \frac{m^i}{{\cal A}^i}- 
 \frac{m^r}{{\cal A}^r} \right) .
\end{eqnarray}
For example, when $m^2/{\cal A}^2 + m^3/{\cal A}^3 -  
m^1/{\cal A}^1 =0$, the massless mode appears in 
$\phi_{z^1}$.
When $m^2/{\cal A}^2 + m^3/{\cal A}^3 -  
m^1/{\cal A}^1 $ is positive (negative), it becomes 
massive (tachyonic).

\section{Couplings including massive modes}

Here, we study couplings including zero-modes and higher modes 
in 4D low-energy effective field theory.
The 3-point couplings among zero-modes 
are computed in \cite{Cremades:2004wa,Antoniadis:2009bg}, 
and higher order couplings among zero-modes 
are studied in \cite{Abe:2009dr}.
First we briefly review on them in section 
\ref{eq:coupling-zero-model}
, and extend to 
3-point and higher order couplings including higher modes 
in section \ref{eq:coupling-massive-model}.
In section \ref{eq:coupling-WL}, 
we also consider the couplings including massive modes due to 
only the Wilson line effect, but not magnetic fluxes.

\subsection{Couplings among zero-modes}
\label{eq:coupling-zero-model}

Here we concentrate on the $T^2$ theory.
We consider the coupling among three zero-modes, 
whose wave functions are given as 
$\psi^{i,q_1m_1}(z+\zeta_1,\tau)$, $\psi^{i,q_2m_2}(z+\zeta_2,\tau)$ 
and $(\psi^{i,q_3m_3}(z+\zeta_3,\tau))^*$.
They have $U(1)$ charges, $q_1$, $q_2$ and $q_3$, respectively, 
and  the magnetic fluxes, $m_1$, $m_2$ and $m_3$ appear 
in their zero-mode equations. 
We use the notation, $N_1=q_1m_1$, $N_2=q_2m_2$ and  $N_3=q_3m_3$.
We assume that  $N_1, N_2,N_3 \neq 0$.
The gauge invariance requires that 
$q_1+q_2=q_3$, $N_1+N_2=N_3$ and $N_1\zeta_1+N_2\zeta_2=N_3\zeta_3$.
Their 3-point coupling in the 4D low-energy effective field 
theory is given by the following integral of wavefunctions,
\begin{eqnarray}
y^{ij\bar k}=\int d^2z \psi^{i,N_1} \psi^{j,N_2} (\psi^{k,N_3})^* ,
\end{eqnarray}
up to the 3-point coupling constant in higher dimensional field theory.
Hereafter, we concentrate on the part given as the overlap integral 
of wavefunctions, omitting the coupling constants in higher dimensions.
For the Yukawa coupling, two of these modes correspond to 
the spinor fields, and the other corresponds to the 
4D scalar field.
The 4D scalar may be originated from the higher dimensional 
vector, e.g. on $T^6$, if the 4D scalar is massless.
At any rate, the wavefunctions are the same among the spinor, 
scalar and vector fields.
Thus, we compute the 3-point and higher order couplings 
without specifying such Lorentz transformation behaviors.
However, note that the Lorentz invariance leads to a certain 
selection rule.

In the computation of the above integral, 
the important property of zero-mode wavefunctions is that 
they satisfy the following relation,
\begin{align}\label{eq:psi-psi-2}
\psi^{i,N_1}(z_1,\tau)\cdot \psi^{j,N_2}(z_2,\tau)&=
\frac{1}{\sqrt{N_1+N_2}}
\sum_{m=1}^{N_1+N_2}
\psi^{i+j+N_1m,~N_1+N_2}(X,\tau )
\notag \\
&\times
\psi^{N_2i-N_1j+N_1N_2m,~N_1N_2(N_1+N_2)} (Y,\tau ) ,
\end{align}
where 
 \begin{equation}
X= \frac{N_1z_1+N_2z_2}{N_1+N_2}, \qquad Y=\frac{z_1-z_2}{N_1+N_2},
\end{equation}
as shown in Appendix \ref{app:prod-wf}
(see also 
\cite{Cremades:2004wa,Antoniadis:2009bg}).

For example, when all of Wilson lines vanish, i.e. 
$z_1=z_2=z$, 
the above expansion 
becomes 
\begin{align}\label{eq:psi-psi-3}
\psi^{i,N_1}(z,\tau)\cdot \psi^{j,N_2}(z,\tau)&=
\left( \frac{2{\rm Im}\tau N_1N_2}{{\cal A}^2(N_1+N_2)}\right)^{1/4}
\sum_{m=1}^{N_1+N_2}
\psi^{i+j+N_1m,N_1+N_2}(z,\tau)
\notag \\
&\times
\vartheta \left[ \begin{matrix} 
\frac{N_2i-N_1j+N_1N_2m}{N_1N_2(N_1+N_2)} \\ 0
\end{matrix} \right]
(0,\tau N_1N_2(N_1+N_2)) .  
\end{align}
Then, by using the orthonormal condition (\ref{eq:orthonormal}), 
the 3-point coupling is obtained as 
\begin{eqnarray}
y^{ij\bar k} = \left( \frac{2{\rm Im}\tau N_1N_2}{{\cal A}^2(N_1+N_2)}\right)^{1/4}
\sum_{m=1}^{N_1+N_2}
\delta_{k,i+j+N_1m}
\cdot
\vartheta \left[ \begin{matrix} 
\frac{N_2i-N_1j+N_1N_2m}{N_1N_2(N_1+N_2)} \\ 0
\end{matrix} \right]
(0,\tau N_1N_2(N_1+N_2)) .   \nonumber \\ 
\end{eqnarray}
There is the selection rule for allowed couplings as 
\begin{eqnarray}\label{eq:selection}
k= i+j, \qquad ({\rm mod}~~N_1).
\end{eqnarray}

Similarly, we can calculate the 3-point coupling for 
non-vanishing Wilson lines.
Its result leads to the 3-point couplings,
\begin{eqnarray}
y^{ij\bar k} &=& \left( \frac{2{\rm Im}\tau N_1N_2}{{\cal A}^2(N_1+N_2)}\right)^{1/4}
\sum_{m=1}^{N_1+N_2}
\delta_{k,i+j+N_1m}  
 e^{i\pi (N_1\zeta_1 {\rm Im}\zeta_1+ N_2 \zeta_2 {\rm Im}\zeta_2 - 
N_3\zeta_3{\rm Im}\zeta_3)/{\rm Im}\tau} \nonumber \\
& & \times \vartheta \left[ \begin{matrix} 
\frac{N_2i-N_1j+N_1N_2m}{N_1N_2(N_1+N_2)} \\ 0
\end{matrix} \right]
(N_1N_2(\zeta_1 - \zeta_2),\tau N_1N_2(N_1+N_2)) .
\end{eqnarray}

Next, we consider the 4-point couplings, 
\begin{eqnarray}
y^{ijk \bar \ell}=\int d^2z \psi^{i,N_1} \psi^{j,N_2} \psi^{k,N_3} (\psi^{\ell,N_4})^* ,
\end{eqnarray}
where the gauge invariance requires $N_1+N_2+N_3=N_4$.
For simplicity we consider the case that all of Wilson lines vanish, 
but it is straightforward to extend to the case with non-vanishing
Wilson lines.
The direct computation is possible by using the relation (\ref{eq:psi-psi-3}).
However, the following calculation is much simpler \cite{Abe:2009dr}.
We write the above integral 
\begin{eqnarray}\label{eq:4-point}
y^{ijk\bar \ell}=\int d^2z d^2z'\psi^{i,N_1}(z) \psi^{j,N_2}(z) \delta^2 (z-z')\psi^{k,N_3}(z') (\psi^{\ell,N_4}(z')))^* ,
\end{eqnarray}
We replace the $\delta$ function by 
\begin{eqnarray}
\delta^2(z-z')= \sum_{s,n} (\psi^{s,N_1+N_2}_n (z))^*\psi^{s,N_1+N_2}_n(z'),
\end{eqnarray}
which is the summation over the complete set corresponding to 
eigenfunctions for the magnetic flux $N_1+N_2$.
This summation  includes higher modes, i.e. $n\neq 0$.
Then we can write 
\begin{eqnarray}
y^{ijk\bar \ell}&=&\sum_{s,n} \left(\int d^2z 
\psi^{i,N_1}(z) \psi^{j,N_2}(z) (\psi^{s,N_1+N_2}_n (z))^* \right) 
\nonumber \\
& & \times \left( \int d^2z' \psi^{s,N_1+N_2}_n(z')
\psi^{k,N_3}(z') (\psi^{\ell,N_4}(z')))^* \right).
\end{eqnarray}
Using the 3-point coupling among zero-modes, the above 
integral can be obtained as 
\begin{eqnarray}\label{eq:4-point-s}
y^{ijk\bar \ell} = \sum_s y^{ij\bar s} y^{sj\bar k}.
\end{eqnarray}
The higher modes $n\neq 0$ do not appear in this 
summation, because only zero-mode modes $n=0$ appear 
in the RHS of Eq.(\ref{eq:psi-psi-3}).

Instead of Eq.(\ref{eq:4-point}), there is another way to 
split the integral, e.g., 
\begin{eqnarray}\label{eq:4-point-2}
y^{ijk\bar \ell}=\int d^2z d^2z'\psi^{i,N_1}(z) \psi^{j,N_2}(z) \delta^2 (z-z')\psi^{k,N_3}(z') (\psi^{\ell,N_4}(z')))^* .
\end{eqnarray}
Then, by replacing the $\delta$ function by 
\begin{eqnarray}
\delta^2(z-z')= \sum_{t,n} (\psi^{t,N_2+N_3}_n (z))^*\psi^{t,N_2+N_3}_n(z'),
\end{eqnarray}
we obtain 
\begin{eqnarray}\label{eq:4-point-t}
y^{ijk\bar \ell} = \sum_t y^{jk\bar t} y^{si\bar \ell}.
\end{eqnarray}
We can show that both of Eqs.(\ref{eq:4-point-s}) and
(\ref{eq:4-point-t}) lead to the same result \cite{Abe:2009dr}.

Similarly, we can calculate another type of the 4-point coupling,
\begin{eqnarray}
y^{ij\bar k \bar \ell}=\int d^2z \psi^{i,N_1} \psi^{j,N_2} (\psi^{k,N_3})^* (\psi^{\ell,N_4})^* ,
\end{eqnarray}
where the gauge invariance requires $N_1 +N_2=N_3+N_4$.
Furthermore, the integrals for 5-point and higher order couplings 
can be carried out in a similar analysis, 
and they are written by the proper summations over products of 
3-point couplings.

\subsection{Couplings including higher modes}
\label{eq:coupling-massive-model}

Here, we study couplings including higher modes.
The relation (\ref{eq:psi-psi-2}) among zero-mode wavefunctions 
plays an important role in 
the computation of the 3-point couplings for zero-modes.
When we operate $(\partial_{z_1}-\frac{\pi N_1}{2\mathrm{Im}\tau} \bar
z_1)^{n_1}
(\partial_{z_2}-\frac{\pi N_2}{2\mathrm{Im}\tau} \bar z_2)^{n_2}$
on the LHS of Eq.~(\ref{eq:psi-psi-2}), we obtain 
\begin{align}\label{eq:psi-psi-mass-1}
&  \bigg(\partial_{z_1}-\frac{\pi N_1}{2\mathrm{Im}\tau}\bar z_1 \bigg)^{n_1}
\bigg(\partial_{z_2}-\frac{\pi N_2}{2\mathrm{Im}\tau}\bar z_2 \bigg)^{n_2}
\psi_0^{i,N_1}(z_1,\tau)\cdot \psi_0^{j,N_2}(z_2,\tau)
\notag \\
&~~~~~~~=\sqrt{  {n_1}!  {n_2}!   \bigg( \frac{\pi N_1}{\mathrm{Im}\tau} \bigg)^{n_1} \bigg( \frac{\pi N_2}{\mathrm{Im}\tau} \bigg)^{n_2}}
\psi_{n_1}^{i,N_1}(z_1,\tau)\cdot \psi_{n_2}^{j,N_2}(z_2,\tau).
\end{align}
On the other hand, when we operate $(\partial_{z_1}-\frac{\pi N_1}{2\mathrm{Im}\tau} \bar
z_1)^{n_1}
(\partial_{z_2}-\frac{\pi N_2}{2\mathrm{Im}\tau} \bar z_2)^{n_2}$
on the RHS of Eq.~(\ref{eq:psi-psi-2}), we obtain 
\begin{align}\label{eq:psi-psi-mass-2}
& \frac{1}{\sqrt{N_1+N_2}}\sum_{m=1}^{N_1+N_2} \sum_{\ell=0}^{n_1} \sum_{s=0}^{n_2}
{}_{n_1}\mathrm{C}_\ell \ {}_{n_2}\mathrm{C}_s
\frac{(-1)^{n_2-s}N_1^\ell N_2^s}{(N_1+N_2)^{n_1+n_2}}
\notag
\\
& \times \bigg(\partial_X
-\frac{\pi }{2\mathrm{Im}\tau}(N_1+N_2)\bar{X} \bigg)^{\ell+s}
\psi^{i+j+N_1m,N_1+N_2}_0(X,\tau)
\notag
\\
& \times \bigg(\partial_Y-\frac{\pi }{2\mathrm{Im}\tau}
N_1N_2(N_1+N_2)\bar{Y}\bigg)^{n_1+n_2-\ell-s}
\psi^{N_2i-N_1j+N_1N_2m,N_1N_2(N_1+N_2)}_0(Y,\tau)
\notag \\
&=\frac{1}{\sqrt{N_1+N_2}}
\sum_{m=1}^{N_1+N_2} \sum_{\ell=0}^{n_1} \sum_{s=0}^{n_2}
{}_{n_1}\mathrm{C}_\ell \ {}_{n_2}\mathrm{C}_s
\frac{(-1)^{n_2-s}N_1^\ell N_2^s}{(N_1+N_2)^{(n_1+n_2)/2}} 
\bigg( \frac{\pi }{\mathrm{Im}\tau} \bigg)^{(n_1+n_2)/2} \notag
\\
&
\times (N_1N_2)^{(n_1+n_2-\ell -s)/2}
\sqrt{ (n_1+n_2-\ell-s)!(\ell+s)!}
\notag
\\
& \times \psi^{i+j+N_1m,N_1+N_2}_{l+s}(X,\tau) \cdot
\psi^{N_2i-N_1j+N_1N_2m,N_1N_2(N_1+N_2)}_{n_1+n_2-\ell-s}(Y,\tau),
\end{align}
by using the derivatives with respect of $X$ and $Y$.
By identifying Eqs.~(\ref{eq:psi-psi-mass-1}) and 
(\ref{eq:psi-psi-mass-2}), the product of higher modes, 
$\psi_{n_1}^{i,N_1}(z_1,\tau)$ and $\psi_{n_2}^{j,N_2}(z_2,\tau)$, 
is expanded as\footnote{A similar relation has been derived 
in twisted tori \cite{BerasaluceGonzalez:2012vb}.} 
\begin{align}
&\psi_{n_1}^{i,N_1}(z_1,\tau)\cdot \psi_{n_2}^{j,N_2}(z_2,\tau)
=
\sum_{m=1}^{N_1+N_2} \sum_{\ell=0}^{n_1} \sum_{s=0}^{n_2}
{}_{n_1}\mathrm{C}_\ell \ {}_{n_2}\mathrm{C}_s
(-1)^{n_2-s} \frac{ N_1^{(n_2+\ell-s)/2}N_2^{(n_1-\ell+s)/2}}
{(N_1+N_2) ^{(n_1+n_2+1)/2}}
\notag
\\
& \times \sqrt{
\frac{(\ell+s)! (n_1+n_2-\ell-s)!}{n_1! n_2!}}
\psi^{i+j+N_1m,N_1+N_2}_{\ell+s}(X,\tau) \cdot
\psi^{N_2i-N_1j+N_1N_2m,N_1N_2(N_1+N_2)}_{n_1+n_2-\ell-s}(Y,\tau).
\end{align}
When we take $z_1 = z + \zeta_1$ and $z_2 = z + \zeta_2$, 
it is found that 
\begin{align}
&\psi_{n_1}^{i,N_1}(z+\zeta_1,\tau)\cdot \psi_{n_2}^{j,N_2}(z+\zeta_2,\tau)
\notag
\\
&
=
\sum_{m=1}^{N_1+N_2} \sum_{\ell=0}^{n_1} \sum_{s=0}^{n_2}
{}_{n_1}\mathrm{C}_\ell \ {}_{n_2}\mathrm{C}_s
(-1)^{n_2-s} \frac{ N_1^{(n_2+\ell-s)/2}N_2^{(n_1-\ell+s)/2}}
{(N_1+N_2) ^{(n_1+n_2+1)/2}}\sqrt{
\frac{(\ell+s)! (n_1+n_2-\ell-s)!}{n_1! n_2!}}
\notag
\\
& \times  
\psi^{i+j+N_1m,N_1+N_2}_{\ell+s}(z+\zeta_3,\tau) \cdot
\psi^{N_2i-N_1j+N_1N_2m,N_1N_2(N_1+N_2)}_{n_1+n_2-\ell-s}(\frac{\zeta_1-\zeta_2}{N_1+N_2},\tau).
\end{align}
Note that the last factor, $\psi^{N_2i-N_1j+N_1N_2m,N_1N_2(N_1+N_2)}_{n_1+n_2-\ell-s}(\frac{\zeta_1-\zeta_2}{N_1+N_2},\tau)$,  is constant.

Using the above relation, we can compute the 3-point coupling,
\begin{align}
&y^{ij\bar k}_{n_1 n_2 n_3} =  \int dzd\bar{z}
\psi_{n_1}^{i,N_1}(z+\zeta_{1},\tau)\cdot \psi_{n_2}^{j,N_2}(z+\zeta_{2},\tau)\cdot
(\psi_{n_3}^{k,N_3}(z+\zeta_{3},\tau))^* ,
\end{align}
in a way to similar to the 3-point coupling among the zero-modes.
The result is obtained as 
\begin{align}
&y^{ij\bar k}_{n_1 n_2 n_3} =
\sum_{m=1}^{N_1+N_2} \sum_{\ell=0}^{n_1} \sum_{s=0}^{n_2}
{}_{n_1}\mathrm{C}_\ell \ {}_{n_2}\mathrm{C}_s
(-1)^{n_2-s}
\sqrt{\frac{ N_1^{n_2+\ell-s}N_2^{n_1-\ell+s}}
{(N_1+N_2) ^{n_1+n_2+1}}
\frac{(\ell+s)! (n_1+n_2-\ell-s)!}{n_1! n_2!}}
\notag
\\
&
\times \psi^{N_2i-N_1j+N_1N_2m,~N_1N_2(N_1+N_2)}_{n_1+n_2-\ell-s}
(\frac{\zeta_1-\zeta_2}{N_1+N_2},\tau)
\delta_{\ell+s,n_3}\delta_{k,i+j+N_1m} \notag
\\
&=
\sum_{\ell=max(0,n_3-n_2)}^{min(n_1,n_3)}
{}_{n_1}\mathrm{C}_\ell \ {}_{n_2}\mathrm{C}_{n_3-\ell}
(-1)^{n_2-n_3-\ell}
\sqrt{\frac{ N_1^{n_2-n_3+2\ell}N_2^{n_1+n_3-2\ell}}
{(N_1+N_2) ^{n_1+n_2+1}}
\frac{n_3! (n_1+n_2-n_3)!}{n_1! n_2!}} \notag \\
& \times \psi^{N_2 k-N_3j,~N_1N_2(N_1+N_2)}_{n_1+n_2-n_3}
(\frac{\zeta_1-\zeta_2}{N_1+N_2},\tau).
\end{align}
There is the selection rule among $i,j$ and $k$, 
which is the same as one for the zero-modes (\ref{eq:selection}).
Thus, the flavor symmetry appearing only in zero-modes 
is still exact even when we take into account the effects due to 
higher modes.
In addition, the following relation, 
\begin{eqnarray}
n_3\leq n_1+n_2 ,
\end{eqnarray}
should be satisfied for the mode numbers, $n_1,n_2$ and $n_3$.
For example, two zero-modes, $n_1=n_2=0$, can couple with 
only the zero mode $n_3 = 0$.
On the other hand, the two zero-modes, $n_1=n_3=0$, can 
couple with higher modes, $n_2 \neq 0$, 
and its coupling is determined by 
\begin{eqnarray}\label{eq:0-0-mass}
\frac{1}{\sqrt{N_1+N_2}}\left(\frac{ N_1}
{N_1+N_2} \right)^{n_2 /2} 
  \psi^{N_2k-N_3j,~N_1N_2N_3}_{n_2}(\frac{\zeta_1-\zeta_2}{N_1+N_2},\tau).
\end{eqnarray}

Similarly, we can compute the 4-point coupling, 
\begin{align}
&y^{ij k \bar \ell}_{n_1 n_2 n_3 n_4} =  \int dzd\bar{z}
\psi_{n_1}^{i,N_1}(z+\zeta_{1},\tau)\cdot
\psi_{n_2}^{j,N_2}(z+\zeta_{2},\tau)\cdot  \psi_{n_3}^{k,N_3}(z+\zeta_{3},\tau)\cdot
(\psi_{n_4}^{\ell,N_4}(z+\zeta_{4},\tau))^* .
\end{align}
We rewrite it as 
\begin{align}\label{eq:massive-split}
&\int d^2zd^2{z'}
\psi_{n_1}^{i,N_1}(z+\zeta_{1},\tau)\cdot
\psi_{n_2}^{j,N_2}(z+\zeta_{2},\tau)\delta^2(z-z') \psi_{n_3}^{k,N_3}(z'+\zeta_{3},\tau)\cdot
(\psi_{n_4}^{\ell,N_4}(z'+\zeta_{4},\tau))^* ,
\end{align}
and replace  the $\delta$ function by 
\begin{equation}\label{eq:massive-delta}
\delta^2(z-z')=\sum_{n,s} \psi_n^{s,N_1+N_2}(z+\zeta,\tau)(\psi_n^{s,N_1+N_2}(z'+\zeta,\tau))^*.
\end{equation}
Then, the 4-point coupling is given as the summation over 
products of 3-point couplings,
\begin{eqnarray}
y^{ij k \bar \ell}_{n_1 n_2 n_3 n_4} = \sum_{n,s}
y^{ij \bar s}_{n_1 n_2 n}~y^{s k \bar \ell}_{n n_3 n_4}.
\end{eqnarray}
Similarly, we can compute other higher order couplings 
by products of the 3-point couplings.

\subsection{Couplings including massive modes only due to 
Wilson lines}
\label{eq:coupling-WL}

In the previous section, we have considered the couplings 
including higher modes under the magnetic flux.
Here, we consider the couplings including massive modes 
only due to the Wilson line.
The wavefunctions of such modes are obtained in 
Eq.~(\ref{eq:massive-WL}).
We compute the following 3-point couplings among 
two zero-modes $\psi^{j,N_1}(z+\zeta_1,\tau)$ 
and $(\psi^{k,N_2}(z+\zeta_2,\tau))^*$ and 
the massive mode $\psi_{n_R,n_I}^{(W)}(z)$, 
where the two zero-modes have non-vanishing magnetic flux, while 
the massive mode has no magnetic flux, but Wilson line.
Here, the gauge invariance requires that 
$N_1 = N_2$ and the Wilson line $\alpha_3$ of the massive mode
$\psi_{n_R,n_I}^{(W)}(z)$ satisfies $N_1 \zeta_1 + \alpha_3 = N_1
\zeta_2$.
Then, the 3-point coupling among these modes 
is given by the following integral, 
\begin{equation}
 y^{j,\bar k}_{(W) n_R n_I} = \int dz d\bar{z}
\psi^{j,N_1}(z+\zeta_1,\tau)
(\psi^{k,N_1}(z+\zeta_{2},\tau))^*
\psi_{n_R,n_I}^{(W)}(z).
\end{equation}
More explicitly, the integral is written by 
\begin{align}
&y^{j,\bar k}_{(W) n_R n_I} \notag \\
&= \frac{\sqrt{2 N_1{\rm Im }\tau}}{{\cal A}^{3/2}}\int dz d\bar{z} \sum_{\ell,n}
\exp
\bigg[\frac{i\pi}{N_1 \mathrm{Im}\tau}\{
(N_1 z+\alpha_{1})\mathrm{Im}(N_1 z+\alpha_{1}) 
\notag \\
&
      -(N_1 \bar{z}+\bar{\alpha}_{2})\mathrm{Im}(N_1 z+\alpha_{2})  \}
+\pi i \bigg(\frac{j}{N_1}+\ell\bigg)^2N_1\tau-\pi i 
\bigg(\frac{k}{N_1}+n\bigg)^2N_1\bar{\tau} 
\notag \\
& 
+2\pi i\bigg\{(N_1 z+\alpha_{1})
\bigg(\frac{j}{N_1}+\ell \bigg)  -(N_1 \bar{z}+\bar{\alpha}_{2})
\bigg(\frac{k}{N_1}+n \bigg)\bigg\}
\notag
\\
&
+\pi i \bigg\{\mathrm{Re}z
\bigg(\frac{\mathrm{Im}\alpha_{3}}{\mathrm{Im}\tau}+2n_R\bigg)
+
\frac{\mathrm{Im}z}{\mathrm{Im}\tau}
\bigg(-\mathrm{Re}\alpha_{3}+2(n_I-n_R \mathrm{Re}\tau)\bigg)\bigg\} \bigg].
\end{align}
The integral over ${\rm Re}z$ imposes $j=k-n_R$ and $\ell =n$.
In addition, the integral over ${\rm Im}z $ is Gaussian-like.
By lengthy computation, it is found that 
\begin{align}
y^{j,\bar k}_{(W) n_R n_I} &=
\frac{1}{\sqrt{{\cal A} }}
\exp
\bigg[
-\frac{\pi}{2N_1 \mathrm{Im}\tau}\{(\mathrm{Im}\alpha_{3}+n_R \mathrm{Im}\tau)^2
+(\mathrm{Re}\alpha_{3}-n_I+n_R\mathrm{Re}\tau)^2)\}
\notag
\\
 &+i \frac{\pi}{N_1{\rm Im} \tau } \bigg( 
{\rm Im} \bar \alpha_2 \alpha_1 + n_R {\rm Im} \bar \tau (\alpha_1 +
\alpha_2)
+   n_R n_I{\rm Im} \tau  -  n_I {\rm Im}(\alpha_1+\alpha_2)
\bigg)
\bigg].
\end{align}
This behaves as a Gaussian function for the Wilson line $\alpha_3$.
Thus, this coupling is suppressed depending on 
the Wilson line as well as  $n_R$ and $n_I$.
The mode with the strongest coupling $|y^{j,\bar k}_{(W) n_R n_I}|$ 
corresponds to the mode with  $n_R=n_I=0$, 
when 
\begin{eqnarray}\label{eq:WL-region}
{}-\frac12 \leq \frac{\rm Im \alpha_3}{\rm Im \tau} \leq \frac12, 
\qquad
{}-\frac12 \leq {\rm Re \alpha_3} \leq \frac12. 
\end{eqnarray}
For other values of $\alpha_3$, another mode with non-vanishing 
$n_R$ and/or $n_I$ would have the strongest coupling. 
For example, for $n_R=n_I=0$, 
we have 
\begin{align}
|y^{j,\bar k}_{(W) n_R=n_I=0}| &=
\frac{1}{\sqrt{{\cal A} }}
\exp
\bigg[
-\frac{\pi|\alpha_{3}|^2}{2N_1 \mathrm{Im}\tau} \bigg].
\end{align}
This coupling is suppressed depending on $|\alpha_{3}|^2$.
For example, when $|\alpha_{3}|^2/2N_1 \mathrm{Im}\tau=1$, 
we obtain  $|y^{j,\bar k}_{(W) n_R=n_I=0}| \approx e^{-\pi} \approx 0.04$. 
The couplings to other modes with $n_R,n_I \neq 0$ are 
much more suppressed for the value of $\alpha_3$, which satisfy 
Eq.~(\ref{eq:WL-region}).

Similarly, we can compute the 3-point coupling among 
two higher modes, $\psi^{j,N_1}_{n_1}(z+\zeta_{1},\tau)$
and $(\psi^{k,N_1}_{n_2}(z+\zeta_{2},\tau))^*$ and the 
massive mode $\psi_{n_R,n_I}^{(W)}(z)$, 
where the first two modes have non-vanishing magnetic flux, while 
the last mode has no magnetic flux, but Wilson line.
We assume that $n_1 \leq n_2$.
Such a coupling is obtained as 
\begin{align}
y^{j,\bar k}_{n_1 n_2 (W) n_R n_I} &= \int dz d\bar{z}
\psi^{j,N_1}_{n_1}(z+\zeta_{1},\tau)
(\psi^{k,N_1}_{n_2}(z+\zeta_{2},\tau))^*
\psi_{n_R,n_I}^{(W)} .
\end{align}
Again, the integral 
over ${\rm Re}z$ imposes $j=k-n_R$ and $\ell =n$.
For the integral over ${\rm Re}z$, we use Eq.~(\ref{eq:integral-I}).
Then, the result is written by 
\begin{align}
y^{j,\bar k}_{n_1 n_2  (W) n_R n_I} &= \frac{
y^{j,\bar k}_{ (W) n_R n_I}}{\sqrt {2^{n_1+n_2}n_2 !}} \sum_{k=0}^{n_1} 2^k \frac{(n_1!)^{3/2}}{(k!)^2(n_1-k)!}
\bigg(\sqrt{2\pi N_1 \mathrm{Im}\tau} \bigg(-\frac{n_R}{N_1}+\frac{\mathrm{Im}\alpha_{3}}
         {N_1 \mathrm{Im}\tau}\bigg)\bigg)^{n_1+n_2-k} .
\end{align}
These couplings include the same suppression factor, 
$y^{j,\bar k}_{ (W) n_R n_I}$.

Higher order couplings can be computed similarly.
When we consider higher order couplings including 
 more modes such as $\psi^{\ell,N}_{n}(\tau,z+\zeta)$,  
we use the technique such as Eqs.(\ref{eq:massive-split}) 
and (\ref{eq:massive-delta}).
When we consider higher order coupling including 
more modes such as $\psi_{n_R,n_I}^{(W)}(z)$, 
we use the property that the product of two wavefunctions 
$\psi_{n_R,n_I}^{(W)}(z)$ and $\psi_{m_R,m_I}^{(W)}(z)$ 
is obtained as $\psi_{n_R+m_R,n_I+m_I}^{(W)}(z)$, and that 
the Wilson line of $\psi_{n_R+m_R,n_I+m_I}^{(W)}(z)$ is 
just the sum of two Wilson lines, which 
$\psi_{n_R,n_I}^{(W)}(z)$ and $\psi_{m_R,m_I}^{(W)}(z)$  have.
Using these, we can compute higher order couplings.

\section{Phenomenological comments}

We have calculated the couplings among zero-modes 
and higher modes.
They have various important implications from the phenomenological 
viewpoints.
Here, we give some comments.

The first example is about the proton decay.
For instance, the proton decay would happen 
through the heavy $X$ boson in the $SU(5)$ GUT model.
It couples with quarks and leptons  by the gauge coupling 
before the gauge symmetry breaking.
This coupling does not change in the 4D GUT theory 
even after the $SU(5)$ group is broken.
However, it can change in extra dimensional models, 
which have been discussed so far.
Let us consider the $SU(5) \times U(1)$ GUT model with 
extra space, $T^2$ or $T^6$.
We introduce non-vanishing magnetic flux $m$ along 
the extra $U(1)$ direction.
Suppose that the ${\bf \bar 5}$ matter field has a charge $q$ 
under the extra $U(1)$ symmetry. 
Before $SU(5)$ breaking, both of the quark and lepton in ${\bf \bar 5}$ are 
quasi-localized at the same place, and their coupling 
to the $X$ boson is given by the gauge coupling.
Then, we assume non-vanishing Wilson line $\alpha$ along the $U(1)_Y$ 
direction in $SU(5)$.
It breaks the $SU(5)$ gauge symmetry, the $X$ boson becomes massive and 
its profile is written by $\psi^{(W)}_{n_R=n_I=0}(z)$ 
in Eq.~(\ref{eq:massive-WL}).
The quark and lepton in ${\bf \bar 5}$ are still massless, but 
their profiles split each other, because of Wilson lines.
In this case, the coupling among the quark, lepton and the 
$X$ heavy boson is not equal to the gauge coupling, but 
it includes the suppression factor, 
$|y^{j,\bar k}_{(W) n_R =n_I=0}|$, as computed in the previous section.
That is important to avoid the fast proton decay.
For example, when $|\alpha_{3}|^2/2N_1 \mathrm{Im}\tau=1$, 
we obtain $|y^{j,\bar k}_{(W) n_R=n_I=0}| \approx e^{-\pi} \approx 0.04$.
Similarly, the coupling of the $X$ boson with quarks and leptons in the 
${\bf 10}$ matter field can be suppressed.
Then, the proton life time would drastically change by 
${\cal O}(10^4-10^5)$.

Similarly, we can study the case that 
$SU(5)$ is broken by the magnetic flux along the $U(1)_Y$ direction, 
and the $X$ boson becomes massive due to 
the magnetic flux.
In this case, the coupling of the $X$ boson with 
quark and lepton has the suppression factor 
given by Eq.(\ref{eq:0-0-mass}).

Let us comment on another example.
The Higgs mode gains its mass by non-vanishing 
Wilson line in certain models.
That is, the Higgs mode corresponds to the open 
string stretching between two parallel D-branes 
( on at least one $T^2$ of $(T^2)^3$) in 
the picture of intersecting D-brane models (see e.g. \cite{Ibanez:2001nd} ).
The Yukawa couplings between this Higgs field and 
massless matter fields include the factor,  
$|y^{j,\bar k}_{(W) n_R =n_I=0}|$.
When the compactification scale is high such as 
the GUT scale and the Planck scale, 
this Wilson line $\alpha$ generating the Higgs mass 
is quite small, $\alpha \ll 1$, and the factor 
$|y^{j,\bar k}_{(W) n_R =n_I=0}|$ is of ${\cal O}(1)$.
It is also important to see the moduli dependence 
of these couplings, that is, 
the dependence of the complex structure $\tau$
and the Wilson line, $\alpha$, which is the open string modulus.
If F terms of complex structure and/or Wilson line moduli 
are non-vanishing, 
the corresponding $A$ terms would appear and 
they are determined by the moduli dependence.
At any rate, it is important to have found that 
the explicit moduli-dependence of these couplings, 
even though their values are of ${\cal O}(1)$.

We would need heavy right-handed neutrino masses 
for the see-saw mechanism.
These masses may be generated by non-perturbative 
effects (see e.g. \cite{Blumenhagen:2006xt,Ibanez:2006da}).
Alternatively, the right-handed neutrino masses 
are generated by Wilson lines.
If such a mass scale is comparable to the compactification 
scale, the couplings of the right-handed neutrino 
with the left-handed neutrino and the Higgs scalar 
would be suppressed.

Finally, we comment on the K\"ahler metric.
The K\"ahler metric of the matter fields is 
diagonal in the flavor basis.
However, they couple with massive modes.
Such couplings may induce off-diagonal entries 
in K\"ahler metric after integrating out massive modes.
Such off-diagonal entries may lead to 
large FCNCs in the gravity-mediated supersymmetry breaking 
scenario \cite{Camara:2011nj}.
However, when those couplings among massles modes and 
massive modes are suppressed,  
off-diagonal entries would be small.
We have shown that the selection rule for 
allowed couplings including higher modes is the same      
 as the one for only zero-modes.
 Thus, if there is a non-Abelian discrete flavor 
 symmetry in massless modes \cite{Abe:2009vi,BerasaluceGonzalez:2012vb}, 
that forbids the off-diagonal entries in 
the K\"ahler metric.
Recall that such a symmetry is not violated by 
effects due to massive modes.

\section{Conclusion and discussion}

We have studied the mass spectrum and wavefunctions 
of zero-modes and higher modes in extra dimensional models with 
magnetic fluxes and Wilson lines.
Furthermore, we have computed 3-point couplings 
and higher order couplings included higher modes 
in the 4D low-energy effective field theory.
These couplings have non-trivial behaviors, because 
wavefunctions of massless and massive modes are 
quite non-trivial.
Using our results, we can write down the 4D low-energy 
effective field theory with the full modes.
Higher modes do not violate the coupling selection rules 
among only zero-modes.
Thus, the flavor symmetry for zero-modes remains 
exact even when we take into account the effects due to 
higher modes.

Our results would be important to phenomenological aspects, 
where couplings between massless and massive modes play 
a role, for example, the proton decay, the Higgs mass term, 
right-handed majorana neutrino mass term, FCNCs, etc.
We will study in detail phenomenological applications of 
our results elsewhere.
Threshold corrections and their moduli dependence 
 after integrating out the massive modes 
would be important.

\subsection*{Acknowledgement}
The authors would like to acknowledge Hiroyuki Abe, Hiroshi Ohki and 
Manabu Sakai for useful discussions. 
The work of T.~K. is supported in part by the Grant-in-Aid for 
the Global COE Program ``The Next Generation of Physics, 
Spun from Universality and Emergence'' from the Ministry of 
Education, Culture, Sports, Science and Technology of Japan.

\appendix

\section{Hermite function}

Here we show properties of the Hermite function, 
$H_n(x)$, which is defined as 
\begin{equation}
H_n(x)=(-1)^n e^{x^2}\frac{d^n}{dx^n}e^{-x^2}.
\end{equation}
Its derivative satisfies 
\begin{align}
& \frac{d}{dx}H_n (x) =2xH_n(x)-H_{n+1}(x),
\\
& \frac{d}{dx}H_n(x)=2nH_{n-1}(x).
\end{align}
The orthonormal conditions are written as 
\begin{align}
\int_{-\infty}^{\infty} H_n(x)H_m(x)e^{-x^2}dx=\delta_{m,n}2^n \sqrt{\pi} n!.
\end{align}

We compute the following integral,
\begin{eqnarray}
I = \int_{-\infty}^{\infty} dx H_n(x+A)H_m(x+B)e^{-(x+A+B)^2},
\end{eqnarray}
for $n\leq m$.
This integral can be calculated as 
\begin{align}\label{eq:integral-I}
I &=\int dx e^{-(x+A)^2}\frac{d^n}{dx^n}(e^{-2B(x+A)-B^2}H_m(x+B))
\notag \\
&=\int dx e^{-(x+A)^2}\sum_{k=0}^n H_m^{(k)}(x+B)(-2B)^{n-k}e^{-2B(x+A)-B^2}{}_{n}\mathrm{C}_k
\notag\\
&=\sum_{k=0}^n 2^k {}_{n}\mathrm{C}_k \  {}_{n}\mathrm{P}_k(-2B)^{n-k}
\int dx e^{-(x+A+B)^2} H_{m-k}^{(k)}(x+B)
\notag\\
&=\sum_{k=0}^n 2^k {}_{n}\mathrm{C}_k \  {}_{n}\mathrm{P}_k(-2B)^{n-k}
\int dx e^{-(x+B)^2}\frac{d^{m-k}}{dx^{m-k}}e^{-2A(x+B)-A^2}
\notag\\
&=\sum_{k=0}^n 2^k {}_{n}\mathrm{C}_k \  {}_{n}\mathrm{P}_k(-2B)^{n-k} (-2A)^{m-k}
\int dx e^{-(x+A+B)^2}
\notag \\
&=\sum_{k=0}^n 2^k {}_{n}\mathrm{C}_k \  {}_{n}\mathrm{P}_k(-2B)^{n-k} (-2A)^{m-k} \sqrt{\pi},
\end{align}
where 
\begin{eqnarray}
H^{(k)}_m(x) = \frac{d^k}{dx^k}H_m(x), \qquad 
{}_{n}\mathrm{C}_k = \frac{n!}{k ! (n-k)!}, \qquad 
{}_{n}\mathrm{P}_k = \frac{n!}{k !}.
\end{eqnarray}
The following integral along the proper path,
\begin{align}
\int_{-\infty}^{\infty} H_n(x+A)H_m(x+B)e^{-(x+A+B+iC)^2},
\end{align}
leads to the same result as the above.

\section{Vector field}
\label{sec:app-vec}

Here, we study the $(4+2)$ dimensional $U(N)$ non-Abelian 
gauge theory(see also \cite{Cremades:2004wa,Conlon:2008qi}).
Its Lagrangian is given as 
\begin{eqnarray}
{\cal L} = -\frac{1}{4g^2}{\rm Tr}\left( F^{MN}F_{MN} \right),
\end{eqnarray}
where 
\begin{eqnarray}
F_{MN} = \partial_M A_N - \partial_N A_M - i[A_M,A_N].
\end{eqnarray}
We compactify the two dimensions on $T^2$ with 
magnetic fluxes along $U(1)$ directions.
We decompose the $U(1)$ parts $B_N$ and 
off-diagonal parts $W_M$, 
\begin{eqnarray}
A_M = B_M+W_M=B^a_MU_a+W_M^{ab}e_{ab},
\end{eqnarray}
with 
\begin{eqnarray}
(U_a)^i_j=\delta_{ai} \delta_{aj}, \qquad 
(e_{ab})_{ij}=\delta_{ai} \delta_{bj}, \quad (a\neq b), 
\end{eqnarray}
where $W^{ab}_{M} = (W^{ba}_M)^*$.
The quadratic terms of $W_M^{ab}$ in the Lagrangian 
are relevant to our study, and these appear 
\begin{eqnarray}
{\cal L}=-\frac{1}{2g^2}{\rm Tr}\left( D_MW_ND^MW^N- D_MW_ND^NW^M -i
G_{MN}[W^M,W^N] \right)+ \cdots,
\end{eqnarray}
where 
\begin{eqnarray}
& & 
G_{MN}=\partial_M B_N - \partial_N B_M,  \\
& & D_M W_N = \partial_M W_N -i [B_N,W_N].
\end{eqnarray}
Here, the ellipsis denotes irrelevant terms.
Furthermore, these terms are written by 
\begin{eqnarray}
{\cal L }&=& \frac{i}{4g^2} \left( G^q_{ij} - G^b_{ij} \right) 
\left( W^{i,ab}W^{j,ba} - W^{j,ab}W^{i,ba} \right) -\frac{1}{2g^2}\bigg[
\left( D_\mu W^{ba}_i D^\mu W^{i,ab} \right)  \nonumber \\
&& +\left( \tilde D_i W^{ba}_j \tilde D^i W^{j, ab} \right) -2
\left( \tilde D_i W_\mu^{ba} D^\mu W^{i,ab} \right) -
\left( \tilde D_i W^{ba}_j \tilde D^j W^{i,ab} \right) \bigg] + \cdots,
\end{eqnarray}
where 
\begin{eqnarray}
\tilde D_i W_j^{ab} = \partial_i W_j^{ab}   -i(B^a_i - B^b_i) W^{ab}_j.
\end{eqnarray}

We expand 
\begin{eqnarray}
W_i^{ab}(x,y) = \sum_n \varphi_{n,i}^{ab}(x) \phi_{n,i}^{ab}(y).
\end{eqnarray}
Then, by imposing the gauge-fixing condition 
$\tilde D^i W_i^{ab}=0$, the equation of motion in the internal 
space is written by 
\begin{eqnarray}\label{eq:vector-mass}
\tilde D_i \tilde D^i \phi_{n,j}^{ab} + 
2i \langle G^{ab,i}_j \rangle \phi_{n,i} = -m^2_n \phi_{j,n}^{ab}.
\end{eqnarray}

\section{Theta function identities and products of 
zero-mode wavefunctions}
\label{app:prod-wf}

Here we study the product of theta functions and 
product of zero-mode wavefunctions (see also 
\cite{Cremades:2004wa,Antoniadis:2009bg}).
The theta function satisfies the following identity,
\begin{eqnarray}\label{eq:theta-identity}
 & & \vartheta \left[\begin{matrix}{\frac{r}{N_1}} \\0  \end{matrix}\right]
(z_1,\tau N_1) \cdot 
\vartheta \left[\begin{matrix} {\frac{s}{N_2}} \\ 0 \end{matrix}
\right]
(z_2,\tau N_2) =
\sum_{m=1}^{N_1+N_2} \vartheta \left[\begin{matrix} 
{\frac{r+s+N_1 m}{N_1+N_2}} \\ 0   
\end{matrix} \right]
(z_1+z_2,\tau (N_1+N_2))\notag \\
& &~~~~~~~~~~~~~~~~~~~~ \times
\vartheta \left[\begin{matrix} 
{\frac{N_2 r-N_1s+N_1 N_2 m}{N_1 N_2(N_1+N_2)}} \\ 0 
\end{matrix} \right]
(z_1 N_2- z_2 N_1,\tau N_1 N_2(N_1+N_2)).
\end{eqnarray}

Using this identity, we can derive the following relations 
among products of zero-mode wavefunctions,
\begin{align}
\psi^{i,N_1}(z_1,\tau)\cdot \psi^{j,N_2}(z_2,\tau)&=
\frac{1}{\sqrt{N_1+N_2}}
\sum_{m=1}^{N_1+N_2}
\psi^{i+j+N_1m,N_1+N_2}\bigg(\frac{N_1z_1+N_2z_2}{N_1+N_2},\tau\bigg)
\notag \\
&\times
\psi^{N_2i-N_1j+N_1N_2m,N_1N_2(N_1+N_2)}\bigg(\frac{z_1-z_2}{N_1+N_2},\tau\bigg) .
\end{align}
Its proof is as follows.
The LHS is explicitly written as 
\begin{align}
\psi^{i,N_1}(z_1,\tau)\cdot \psi^{j,N_2}(z_2,\tau)&= \left( \frac{2
    {\rm Im} \tau \sqrt{N_1N_2}}{{\cal A}^2}\right)^{1/2} \exp
\left[\frac{i\pi}{\mathrm{Im}\tau}(N_1z_1\mathrm{Im}z_1+
N_2z_2\mathrm{Im}z_2)\right]
\notag \\
& \times \vartheta \left[ \begin{matrix} {\frac{i}{N_1}} \\  0 
\end{matrix} \right] 
(N_1z_1,N_1\tau) \cdot \vartheta 
\left[ \begin{matrix} {\frac{j}{N_2}} \\0 \end{matrix} \right] (N_2z_2,N_2\tau),
\end{align}
and it can be rewritten by use of Eq.~(\ref{eq:theta-identity}) as 
\begin{align}\label{eq:psi-psi}
\psi^{i,N_1}(z_1,\tau)\cdot \psi^{j,N_2}(z_2,\tau)&  =\left( \frac{2
    {\rm Im} \tau \sqrt{N_1N_2}}{{\cal A}^2}\right)^{1/2} \exp
\bigg[\frac{i\pi}{\mathrm{Im}\tau}(N_1z_1\mathrm{Im}z_1+N_2z_2\mathrm{Im}z_2)\bigg]
\notag \\
& \times \sum_{m=1}^{N_1+N_2} \vartheta \left[ \begin{matrix} 
{\frac{i+j+N_1 m}{N_1+N_2}} \\ 0 \end{matrix} \right] 
(N_1z_1+N_2z_2,\tau (N_1+N_2))\notag \\
&\times 
\vartheta \left[ \begin{matrix} 
{\frac{N_2 i-N_1j+N_1 N_2 m}{N_1
    N_2(N_1+N_2)}} \\ 0 \end{matrix} \right] 
(N_1N_2(z_1-z_2),\tau N_1 N_2(N_1+N_2)).
\end{align}
Since the exponent part is written as 
\begin{align}
& \frac{i\pi}{\mathrm{Im}\tau}(N_1z_1\mathrm{Im}z_1+N_2z_2\mathrm{Im}z_2)
= \frac{i\pi}{\mathrm{Im}\tau}\bigg\{(N_1z_1+N_2z_2)
\mathrm{Im}\bigg(\frac{N_1z_1+N_2z_2}{N_1+N_2}\bigg) 
\notag \\ &~~~~~~~~~~~~~~~~~~~~~~~~~ +
N_1N_2(z_1-z_2)\mathrm{Im}\frac{z_1-z_2}{N_1+N_2}\bigg\},
\end{align}
The RHS in Eq.~(\ref{eq:psi-psi}) is the summation over 
the products of wavefunctions, \break  
$\psi^{i+j+N_1m,N_1+N_2}(\frac{N_1z_1+N_2z_2}{N_1+N_2},\tau)$ and 
$\psi^{N_2i-N_1j+N_1N_2m,N_1N_2(N_1+N_2)}(\frac{z_1-z_2}{N_1+N_2},\tau) $.

\end{document}